\documentclass[usenatbib]{mn2e}
\usepackage[dvips]{graphicx}
\usepackage{natbib}
\usepackage{aas_macros}
\bibpunct[, ]{(}{)}{,}{a}{}{,}
\newcommand{\mmbox}[1]{\mbox{$#1$}}
\newcommand{\dif}{\mathrm{d}}
\newcommand{\diff}[2]{\frac{\dif#1}{\dif#2}}
\newcommand{\fkomma}{\,\mbox{,}}\newcommand{\fpunkt}{\,\mbox{.}}
\newcommand{\gw}[1]{{\left\langle#1\right\rangle}_W}
\newcommand{\pct}{\,\%}
\newcommand{\unit}[1]{\mbox{$\,\mathrm{#1}$}}
\newcommand{\rperi}{r_0}
\newcommand{\rkrit}{\rperi{}_\mathrm{\!,crit}}
\newcommand{\msun}{M_{\sun}}\newcommand{\mterr}{M_\oplus}

\newcommand{\tmsun}{$\msun$}
\newcommand{\mdisk}{M_\mathrm{d}}
\newcommand{\rmin}{r_\mathrm{min}}\newcommand{\rmax}{r_\mathrm{max}}
\newcommand{\mstar}{m}\newcommand{\disk}{\mathrm{d}}
\newcommand{\mcl}{M_\mathrm{cl}}\newcommand{\ncl}{N_\mathrm{cl}}
\newcommand{\rhm}{R_\mathrm{0.5}}\newcommand{\rpl}{R_\mathrm{Pl}}
\newcommand{\nce}{n_\mathrm{enc}}\newcommand{\pce}{P_\mathrm{enc}}
\newcommand{\inc}{\imath}\newcommand{\tilt}{\beta}
\newcommand{\qmin}{Q_\mathrm{min}}

\begin{document}
%
%
\title{Induced planet formation in stellar clusters -- a parameter study of star-disk encounters}
\author[Thies, Kroupa, Theis]{Ingo Thies$^{1,2,5}$, Pavel Kroupa$^{1,2,3,5}$,
Christian Theis$^{1,4}$\\
$^1$Argelander Institut f\"ur Astronomie (Sternwarte), Universit\"at Bonn, Auf dem H\"ugel 71, D-53121 Bonn, Germany\\
$^2$Institut f\"ur Theoretische Physik und Astrophysik der
Universit\"at Kiel, Leibnizstra{\ss}e 15, D-24098 Kiel, Germany\\
$^3$Heisenberg Fellow\\
$^4$Institut f\"ur Astronomie der Universit\"at Wien, T\"urkenschanzstra{\ss}e 17, A-1180 Wien, Austria\\
$^5$The Rhine Stellar Dynamical Network}
\maketitle
\begin{abstract}
We present a parameter study of the possibility of tidally triggered disk instability.
Using a restricted $N$-body model which allows for a survey of an extended parameter space,
we show that a passing dwarf star with
a mass between 0.1 and 1~\tmsun\ can probably induce gravitational instabilities
in the pre-planetary solar disk for
prograde passages with minimum separations below 80--170~AU for isothermal or adiabatic disks.
Inclined and retrograde encounters lead to similar results but require slightly
closer passages. Such encounter distances are quite likely in young moderately massive star clusters (\citealt{scally&clarke:2001};
\citealt{bonnell&al:2001}).
The induced gravitational instabilities may lead to enhanced planetesimal
formation in the outer regions of the protoplanetary disk, and
could therefore be relevant for the existence of Uranus and Neptune, whose formation timescale
of about 100~Myr \citep{2000prpl.conf.1081W} is inconsistent with the disk lifetimes of about
a few Myr according to observational data by \citet{haisch&lada&lada:2001}.
The relatively small gas/solid ratio in Uranus and Neptune can be matched if the perturbing
fly-by occurred after early gas depletion of the solar system, i.e. when the solar 
system was older than about $5$~Myr.

We also confirm earlier results by \citet{heller:1993} that the observed 7 degree
tilt of the solar equatorial plane relative to the ecliptic plane could be the
consequence of such a close encounter.
\end{abstract}

\section{Introduction}
\label{sec:intro}
\subsection{Limits of the coagulation model}
In the classical model of planet formation dust grains collide and stick together forming
larger grains and clumps. These clumps also collide with each other forming progressively
larger planetesimal bodies. At some point the mass of a planetesimal
provides sufficient gravitation to attract surrounding particles---the accretion process
begins.
While the coagulation process takes several Myr for Jupiter and Saturn \citep*{ThoDunLev2002},
it takes at least 100~Myr for Neptune
\citep{hollenbach&al:2000} and probably even longer for objects in the Edgeworth-Kuiper Belt.
The timescale $T_\mathrm{coag}$ of the coagulation phase is
mainly determined by the density $\varrho$ of the surrounding protoplanetary dust. It further
depends on the epicyclic frequency $\kappa$ since the epicyclic motion
is an important factor for the collision rate of planetesimals.
$\kappa$ is given by
\begin{equation}\label{epicycle}
\kappa^2=\left(r\diff{\Omega^2}{r}+4\Omega^2\right)\,,
\end{equation}
where $\Omega$ is the orbital angular frequency and $r$ the radius of the particle orbit, i.e.\ the distance from the central star.
The dust density $\varrho$ is approximately proportional to the surface density $\Sigma$, which is
expected to fit a $r^{-3/2}$ law \citep{1977Ap&SS..51..153W, 2004ApJ...609.1045M}.
Thus for low mass disks
\begin{equation}
\kappa\approx\Omega\approx\Omega_\mathrm{Kepler}\propto r^{-3/2}\fkomma
\end{equation}
where $\Omega_\mathrm{Kepler}$ is the Keplerian frequency.
The relationship between $T_\mathrm{coag}$ and $r$ in a gaseous disk containing fine dust
with negligible gravitational interaction between the dust grains can be written as
(\citealt{safronov1969}; see also \citealt{RiceArmitage2003})
\begin{equation}\label{tcscale}
T_\mathrm{coag} \propto \frac{1}{\Omega\Sigma}\propto r^3\fpunkt
\end{equation}
The relation may be somewhat less steep due to the smaller end masses of the outer planets and
gravitational focusing for larger dust particles.
This is in agreement with \citet{Polletal1996} and
\citet{KokuboIda2000} who found \mmbox{T_\mathrm{form}\propto r^\eta} with \mmbox{eta=2}
and \mmbox{\eta\approx2.6}, respectively.
Assuming a typical formation timescale for Jupiter's core of
3~Myr (\mmbox{r_\mathrm{Jup}=5.2\unit{AU}}) this relation leads to prohibitively long formation
timescales up to 600~Myr. Even the conservative $r^2$ law yields timescales of 40~Myr for Uranus
and 100~Myr for Neptune.
Thus we find good agreement of the formation-time scaling given by Eq. \ref{tcscale} with the
more detailed investigations discussed at the beginning of this section.

Recent work by \citet{2004AJ....128.1348R}
predicts shorter formation times in the order of $\sim10^7$~yr for the core of an ice giant at
30~AU from the Sun. Although this is only one-sixth of the conventional value, it is also near
the disk lifetime discussed below and even beyond, if the accretion time of the envelope and
the precursory
depletion of large planetesimals ($>$0.1--1~km, \citealt{2004AJ....128.1348R}) is taken into
account. \citet{2004ApJ...614..497G} also predict similarly short formation timescales but
require a six times higher surface density (compared to the minimum mass solar nebula, MMSN; see
\citealt{1977Ap&SS..51..153W} and \citealt{1981PThPS..70...35H}) in
the region of Uranus and Neptune to allow them to collect most of their final mass until the
end of oligarchy. This would also imply a significantly higher total disk mass.

The lifetime of the protoplanetary disk is most probably less than 10~Myr.
Based on JHKL-excess observations \citet*{haisch&lada&lada:2001} found lifetimes less than 6~Myr
before the disk material is dispersed by the stellar wind, photoevaporation and gravitational
scattering by protoplanets. The masses of observed disks range from 0.003 to 0.3~\tmsun, with the
bulk below 0.03~\tmsun \citep{Natta04}. This is comparable to the MMSN which is between
0.01 and 0.07~\tmsun.
Thus, unless an unusual high disk mass or surface density profile is assumed, there is
a serious problem to explain the existence of Uranus and Neptune.
There have been several theoretical attempts to solve this contradiction:
\begin{enumerate}
  \item The coagulation and accretion process could be much faster
  than predicted by current models. But it seems unlikely that the
  models overestimate the formation times by more than an order of magnitude.
  Even the fast accretion scenario by \citet{2004AJ....128.1348R} can hardly
  solve this inconsistency.
  \item Protoplanetary disks could survive much longer than expected
  from observations like those by \citet{haisch&lada&lada:2001},
  since only the dust component of a disk can be detected by current
  observational methods, but no solid planetesimals. But as Uranus and
  Neptune mainly consist of solid material and only about 10\% of gas
  \citep{2000prpl.conf.1081W},
  this does not seem to be a satisfactory attempt either.
  \item Uranus and Neptune may have formed much closer to the Sun in
  the birth region of Jupiter and Saturn and ejected from their
  initial orbits after reaching a certain mass
  \citep{ThoDunLev2002}. With this scenario, however, it is difficult
  to understand the almost circular actual orbits of Neptune and Uranus.
\end{enumerate}

\subsection{Gravitational instabilities}
Fragmentation due to gravitational instabilities (GIs) is another possible
solution to this problem. The dynamical free-fall timescale of a GI in a cloud
with a density excess $\varrho_\mathrm{exc}$ (relative to the ambient
medium) can be estimated as
\begin{equation}\label{freefall}T_\mathrm{ff}=\sqrt{\frac{3\pi}{32\,G\varrho_\mathrm{exc}}}
\end{equation}
\citep{binney&tremaine:1987}, where $G$ is the gravitational constant. For an over-density
\mmbox{\varrho_\mathrm{exc}=10^{-12}\unit{g\,cm^{-3}}} (which would result by doubling the
ambient density locally in the disk at 30~AU, the today's orbital radius of Neptune)
$T_\mathrm{ff}\approx100\unit{yr}$, which is six orders of magnitude shorter than the
classical formation timescale and can therefore be neglected compared to the subsequent
accretion phase.

There are two major ways for the
development of GIs. A very massive disk of more than 0.1~\tmsun\ becomes
unstable to axisymmetric perturbations according to the disk stability criterion
$Q$ developed by \citet{toomre:1964},\
\begin{equation}\label{qtoomre}Q \equiv \frac{c_s\kappa}{\pi G\Sigma}\,,
\end{equation}
where $c_s$ is the speed of sound in the gas, $\kappa$ the epicyclic frequency and
$\Sigma$ the surface density.

A disk can become unstable if it continues to
accrete material from the protostellar cloud, so that $Q$ becomes
less than 1.
For non-axisymmetric perturbations in gaseous disks, the limit is higher
at $Q\simeq\sqrt{3} \approx 1.7$ \citep*{1997AstL...23..483P}.

Another way towards local instabilities is via tidal perturbations.
With this contribution we argue that also initially stable disks can become locally
unstable if an external perturbation on the surface density of the disk causes
$Q$ to fall below 1 locally, so that fragmentation only occurs in such 
regions. For example and as shown below, a star with mass \mmbox{m=0.5\,\msun}
on a hyperbolic orbit with eccentricity $e=1.5$ and pericentre distance $\rperi=120\unit{AU}$ may
cause a local Toomre instability within 500 years. 
Such encounters are expected to occur
with a realistic probability within the first 6~Myr in the Orion Nebula Cluster
(\citealt{scally&clarke:2001}; \citealt{bonnell&al:2001}).

The tilt of the solar rotational axis
against the normal direction of the ecliptic plane of about 7\degr\
\citep{eggers&al:1997,heller:1993} is interesting in this respect.
We confirm the results presented by \citet{heller:1993}, who
used smoothed particle hydrodynamics (SPH) instead of our restricted $N$-body calculation,
that such an inclined encounter leads to the observed tilt.

In this paper we study if gravitational instabilities, as a possible originator of
planet formation, can be triggered
due to stellar fly-bys in a parameter survey of the stellar mass,
encounter distance and eccentricity, $\mstar$, $\rperi$ and $e$, respectively.
The aim of this survey is
to identify those regions in the $\mstar$-$\rperi$-$e$ parameter space where growing
local instabilities can be triggered.
Due to the extended parameter space such a study can only be performed by fast
numerical methods. Though detailed hydrodynamical calculations would be
preferable for such models, they are too CPU expensive for scanning a large set of
parameter space. Therefore, we applied a simple numerical ansatz based on the fast
restricted N-body method as was already used by \citet{ida&larwood&burkert:2000}.
In the future we will use our present survey 
to perform hydrodynamical calculations for a few interesting $m$, $\rperi$ and $e$
in order to investigate the scenario of tidally triggered GIs in more detail.

In \S~\ref{sec:model} we describe our numerical
model. \S~\ref{sec:sim} describes the general effects of star-star encounters on circumstellar
disks for an individual example.
\S~\ref{sec:parscan} gives a plan of the parameter study and shows
the results for coplanar and non-coplanar encounters. Orbits of
protoplanet candidates and the tilt of the circumstellar disk due to inclined fly-bys
are also described.
In \S~\ref{sec:discussion} the results are discussed and the conclusions are presented in
\S~\ref{sec:concl}.

\subsection{Encounter probability}
\label{ssec:probenc}

Tidally-induced gravitational instabilities can only be a viable mechanism if encounters
are of sufficient likelihood. To estimate the encounter probability we need to consider
a model of a young cluster.

Assuming a Plummer star-cluster model with given half-mass radius
\mmbox{\rhm=\rpl/\sqrt{2^{2/3}-1}\approx1.305\,\rpl} ($\rpl$ is the Plummer-radius), mass $\mcl$
and number of stars $\ncl$, we can calculate the crossing time $T_\mathrm{cr}$ from the
characteristic velocity dispersion and compute the expected number
of encounters $\nce$ within a given impact parameter $b$ and therefore the mean time between
encounters. This gives us a useful guide-line as to how likely the interesting encounters may be.

\begin{figure}\begin{center}
\includegraphics[width=0.47\textwidth]{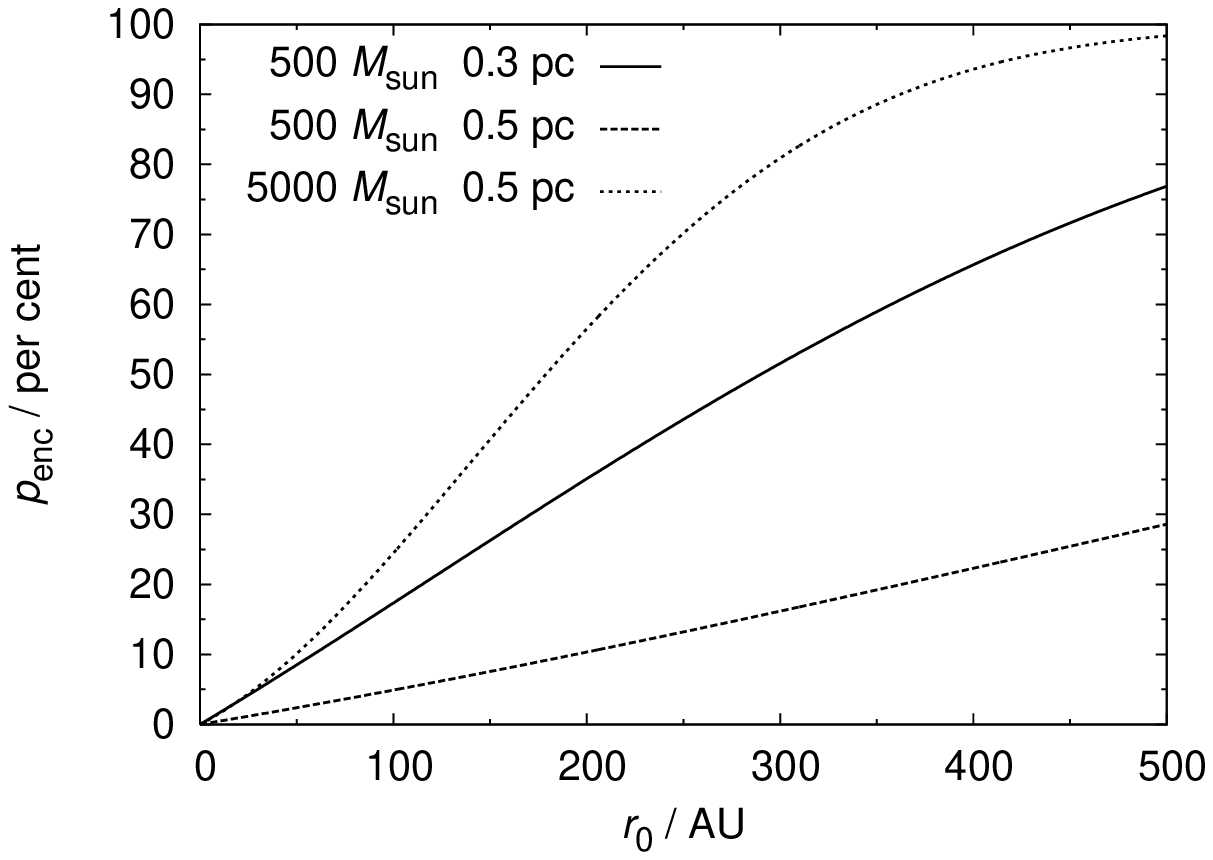}
\caption{\label{penc}Probability $\pce$ of close encounters within \mmbox{T_\disk=6\unit{Myr}}
as a function of the
minimum encounter distance $\rperi$ (Eq. \ref{rvonb}) for three different Plummer
clusters: A young open cluster of high density (solid curve), a more expanded
cluster of the same mass (thick dotted curve) and a young ONC-type
cluster \protect\citep[thin dotted curve,][]{2001MNRAS.321..699K}.
$\pce$ increases with increasing cluster mass $\mcl$ and
decreasing half-mass radius.}
\end{center}\end{figure}

The relationship between $b$, the fly-by eccentricity $e$ and the pericentre distance $\rperi$ is
\begin{equation}\label{rvonb}\rperi=\sqrt{\frac{e-1}{e+1}}\cdot b\fpunkt\end{equation}
The one-dimensional characteristic velocity dispersion $\sigma_\mathrm{1D}$ in the Plummer model, derived
from the virial theorem, is given by
\begin{equation}\sigma_\mathrm{1D}^2=\frac{\pi}{32}\,\frac{G\mcl}{\rpl}=\frac{\pi\sqrt{2^{2/3}-1}}{32}\,\frac{G\mcl}{\rpl}
\fkomma\end{equation}
and the crossing time is
\begin{equation}T_\mathrm{cr}\equiv\frac{2\,\rhm}{\sigma_\mathrm{1D}}=\left(\frac{128\rpl^3}{\pi\left(2^{2/3}-1\right)\cdot G\mcl}\right)
\fpunkt\end{equation}
This leads to the mean time between encounters
\begin{equation}T_\mathrm{enc}=\frac{T_\mathrm{cr}}{\nce}=\frac{T_\mathrm{cr}}{\ncl}\frac{\rhm^2}{b^2}\fpunkt\end{equation}
The probability $\pce$ for at least one event within the disk lifetime $T_\disk$ can be derived from the
Poisson-probability for $x$ encounters,
\begin{equation}\pce(x)=\frac{\nu^x}{x!}\exp\left(-\nu\right)\ ,\quad\mbox{where}\quad\nu=\frac{T_\disk}{T_\mathrm{enc}}\fpunkt\end{equation}
For at least one event,
\begin{equation}\pce(x\ge1)=1-\pce(0)=1-\exp\left(-\frac{T_\disk}{T_\mathrm{enc}}\right)\fpunkt
\end{equation}
For a young cluster with \mmbox{\mcl=500\,\msun}, \mmbox{\rhm=0.3\unit{pc}} and with a
mean stellar mass of 0.5~\tmsun\ we get \mmbox{10\pct<P_\mathrm{enc}<32\pct} within
\mmbox{T_\disk=6\unit{Myr}} for
\mmbox{60\unit{AU}<\rperi<200\unit{AU}}, as shown in Fig \ref{penc}. This means that such
close encounters are indeed likely in young stellar clusters.
The chances are even greater, up to 50\pct, for a cluster of 5000~\tmsun\ and
\mmbox{\rhm=0.5\unit{pc}}, similar to the Orion Nebula Cluster (ONC)
\citep*{2001MNRAS.321..699K}.
\citet{scally&clarke:2001}
propose an encounter probability of only 4\pct\ for encounters closer than 100~AU in the ONC at its
present density and age. This corresponds to a probability of 9\pct\ for \mmbox{\rperi<150\,\mathrm{AU}},
and is still a reasonably high likelihood. We note that the probabilities for encounters
in a younger and denser ONC would be accordingly higher.

\section{The model}
\label{sec:model}
 
We investigate a large region of parameter space in order to study in which fraction of it tidally
induced GI may be relevant. Our parameter space is defined by the distance of
closest approach $\rperi$, the eccentricity of the relative orbit $e$ and the mass of the 
perturbing star $\mstar$. While an exact description of gravitational instabilities requires a 
3d-stellar-hydrodynamical calculation, we apply a simpler, but numerically much faster
restricted N-body approach. This ansatz is valid as long as the self-gravity of the disk,
the pressure forces or the inelastic collisions between the constituents of the
disk can be neglected. For example, simulations of encounters between galaxies have demonstrated
the validity of this approach \citep{1972ApJ...178..623T}.
Because 3d-hydrodynamical calculations are by far 
too CPU intensive at the moment, such simplified dynamical modelling is the only
way to scan the parameter space in detail. Of course, we cannot follow the dynamical
evolution of substructures when their self-gravity becomes important or even dominant.
However, we can identify regions which are prone to such instabilities.

 In our computations the disk is represented by test particles on initially circular and 
coplanar orbits. There is no explicit
interaction (self-gravity, coagulation, pressure forces) between the particles themselves
and also no gravitational feedback of the disk to the Sun or the perturbing star. 
We assume both stars to move on hyperbolic orbits. This should be typical for young star clusters.
Our numerical method requires low computational effort. 
Therefore, it allows for the requested in-depth parameter study of the main orbital
parameters $\mstar$, $e$ and $\rperi$ providing an estimate
of the order of magnitude of the perturbation and its probable
effects on disk stability. We expect this approach to give useful insights into those regions
of the encounter-parameter space where the physics will be of interest for triggered GI,
because the primary mechanism driving the fluctuations in disk density
within the first 500 years
comes from the changes
of the overall potential due to the relative orbit of the two stars.

\subsection{Disk properties}
\label{ssec:diskprop}
In our model we suppose an initially flat disk with an outer radius
\mmbox{\rmax=100\unit{AU}}, an inner border
\mmbox{\rmin=3\unit{AU}} and a total mass  \mmbox{\mdisk=0.073\,\msun},
which is close to the Minimum Mass Solar Nebula.
The surface density $\Sigma$ is then given by
\begin{equation}\label{sigmaprofile}
   \Sigma(r)=6300\unit{g\,cm^{-2}}\cdot\left(\frac{r}{\mathrm{AU}}\right)^{-3/2}\, .
\end{equation}
For later analysis (\S~\ref{ssec:parscantoomre}) the disk is assumed to contain a
dust fraction of the order of 1 per cent \citep{Natta04}. Thus the surface density
of the dust component is approximately 1/100th of the total $\Sigma$,
\mmbox{\Sigma_\mathrm{dust}(r)\approx0.01\Sigma(r)}.
The density profile is needed for the calculation of the local
Toomre parameter $Q$ which we use as an indicator
of local gravitational instability. As shown in Eq. \ref{qtoomre} $Q$ depends on the speed of sound
in the gas fraction of the disk, which mainly depends on the radial temperature
profile. Following a recommendation by \citet{2004AJ....128.1348R} we use
\begin{equation}c_s\approx1200\unit{m\,s^{-1}}\left(\frac{r}{\mathrm{AU}}\right)^{-0.25}\fpunkt
\end{equation}
Thus, the initial $Q$ is approximately given by
\begin{equation}Q_0\approx18\left(\frac{r}{\mathrm{AU}}\right)^{-0.25}\fpunkt
\end{equation} 
The changes of $Q$ also depend on the adopted polytropic index $\gamma$, which
is given with respect to the cooling efficiency of the disk. Large cooling rates prevent
tidally compressed clouds from significant heating, i.e.\ we have an isothermal gas,
\mmbox{\gamma=1}. If the cooling is negligible the gas is compressed adiabatically.
For a di-atomic gas \mmbox{\gamma=1.4} which is a good approximation for the protoplanetary
gas since it mainly consists of molecular hydrogen.
Since \mmbox{p\propto\rho^\gamma} and \mmbox{c_s\propto\sqrt{p/\rho}} and
\mmbox{\Sigma\propto\rho}, the dependence
of $Q$ on the dynamic compression factor $\mathcal{K}$ (\mmbox{\equiv\Sigma/\Sigma_0}) is given by
\begin{equation}Q(\mathcal{K})=Q_0\mathcal{K}^{\frac{\gamma-3}{2\gamma}}\fpunkt\end{equation}

The relations above are valid for gaseous disks. For disks that have already been depleted of
much of their gas, e.g. either through photo-evaporation by their central star or from a nearby
O star, the GIs may develop in the dust component in a similar way. The stability of a dust particle
disk is then given by
\begin{equation}
\label{qdust}Q_\mathrm{dust} = \frac{\sigma_\mathrm{dust}\kappa}{\pi G\Sigma_\mathrm{dust}}\,,
\end{equation}
where $\sigma_\mathrm{dust}$ is the velocity dispersion of the dust particles. 
If the disk particles and gas molecules have reached thermodynamical energy equipartition,
$\sigma_\mathrm{dust}$ can be estimated via the ratio
of the dust particle mass $m_p$ compared to the molecular mass $\mu$ of the gas:
\begin{equation}
\sigma_\mathrm{dust}\sim c_s\cdot\sqrt{\frac{\mu}{m_p}}\fpunkt
\end{equation}
For micron-sized or larger particles this is negligible compared to other causes of
velocity dispersion (e.g. gas turbulences which excite the dust component via gas drag).
However, it is reasonable to assume the dust component to be dynamically much cooler for
particles of a certain size than the gas component. The dynamical friction
damps the velocity dispersion of the dust grains over a long time but the grains remain
inert to short-time gas perturbations and turbulence.
\citet{BFetal2005} show that intermediate sized dust grains (between about 1 mm and 1 m)
tend so settle efficiently into a thin dust layer in the mid-plane of the disk.
If we assume that the disk is already at an intermediate age,
i.e. a few Myr, so that the dust has partly coagulated to at least
centimetre-sized grains, a significant fraction of the dust will be concentrated in or close
to the mid-plane with a low velocity dispersion. A partial depletion of the gas component will
lower the particle size required for most efficient settling.
Therefore we expect \mmbox{Q_\mathrm{dust}<Q} and thus a higher likelihood
for GIs within the dust layer once the gas has been largely depleted.
However, the extension of the simulation to this case has a more qualitative than quantitative
character and needs to be ascertained with hydrodynamical models.
The role of the dust is discussed in more detail in \S~\ref{sec:discussion}.

\subsection{Numerical methods}
\label{ssec:numeric}
Because we neglect particle-particle interactions and the disk-to-star influence,
the gravitational acceleration for each test particle depends only on the position of the
particle and both stars at a given time $t$, whereas the star-star encounter is reduced
to a two-body problem. This allows to split the motion of
the bodies into two parts:
\begin{enumerate}
      \item The hyperbolic (or parabolic) encounter of the stars, and
      \item the motion of the individual particles in the time-dependent gravitational
            field of both stars.
\end{enumerate}
For each particle the equations of motion are solved numerically. In each time step
the acceleration is computed from the time-dependent stellar positions.

The time-dependent positions of the two stars are given by a (semi-)analytical calculation
of the orbits. 
This is reduced to finding the roots of an ordinary equation. In case
of parabolic orbits it can be solved analytically, whereas in the case
of hyperbolic orbits a numerical solution is required.
For the latter we use a Newton-Raphson method. In order to reduce the computational 
costs, we interpolate the orbits by a cubic spline making use of pre-tabulated orbital
positions. The equation of motion of the test particles is solved by a Bulirsch-Stoer
method with adaptive step size. To avoid singularities, a gravitational softening 
with a length of 0.001~AU has been applied.

To obtain the local Toomre parameter the surface density is measured by ``probe particles''
(probes), which sum-up within their probe-volume the masses assigned to the test particles.
The method is described in more detail in Appendix \ref{sec:denseval}.

\section{Simulation of individual encounters}
\label{sec:sim}
To demonstrate the general effects of star-star encounters we simulated the fly-by of a star
with 0.5~\tmsun with \mmbox{\rperi=120\unit{AU}} pericentre distance and an eccentricity of
\mmbox{e=1.5}. For comparison we reran the fly-by with a larger encounter distance
of \mmbox{\rperi=150\unit{AU}}.

\subsection{Initial conditions}
\label{ssec:initial}
The disk particles are initially set on circular coplanar orbits around the Sun. Depending on the
kind of simulation, the particle positions are set up as a 
polar grid arrangement (regular distribution) or as a
random distribution preserving the mean density distribution.
For density evaluations
the regular distribution yields less noisy numerical results. On the other hand, the
calculation of the mean angular momentum (for the estimate of the tilt of the
disk) requires the radial mass distribution to be as smooth as possible. Thus, we
used a random reshuffling of the initial particle positions for calculating the tilt
in \S~\ref{ssec:tilt}, employing the Mersenne Twister algorithm \citep{mersennetwister}.

The initial distance of the Sun and the perturber is set larger than 1000~AU to prevent any
significant tidal effects before the encounter. For the same reason, the simulation ends after
the distance has reached at least the initial value.
For higher resolution, the computed
population of 250,000 particles and 1000 probes is limited to the area between 25 and 30~AU.
Such a selection for increasing the spatial resolution is possible because there are no interactions 
between the particles. 
The absence of further interactions also allows for a simple scaling to other disk radii and
densities, so the results can be scaled up or down for the orbits of Uranus or Pluto and the larger Kuiper belt objects.
In a special run, which is described in \S~\ref{ssec:exorbs}, we extend the simulated
area to 100~AU, the known region of the Edgeworth-Kuiper-Belt. We also extended the disk area
to 50~AU but reduced the particle number to 10,000 for the plots shown in
Figs. \ref{six_120au}.

\subsection{Passage of a 0.5~\tmsun\ star at 120~AU}
\label{ssec:refmodel}
\begin{figure}\begin{center}
\includegraphics[width=0.47\textwidth]{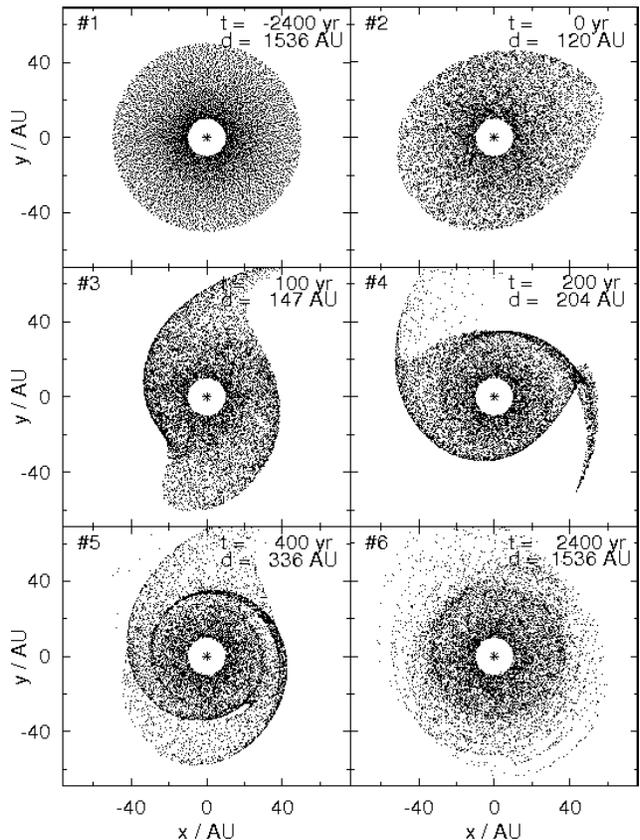}
\caption{\label{six_120au}Disk perturbations during hyperbolic fly-by ($e=1.5$)
of a star with 0.5~\tmsun\ at \mmbox{\rperi=120\unit{AU}}.
Only the regions within an initial radius of 50~AU are shown.}
\end{center}\end{figure}

\begin{figure}\begin{center}
\includegraphics[width=0.47\textwidth]{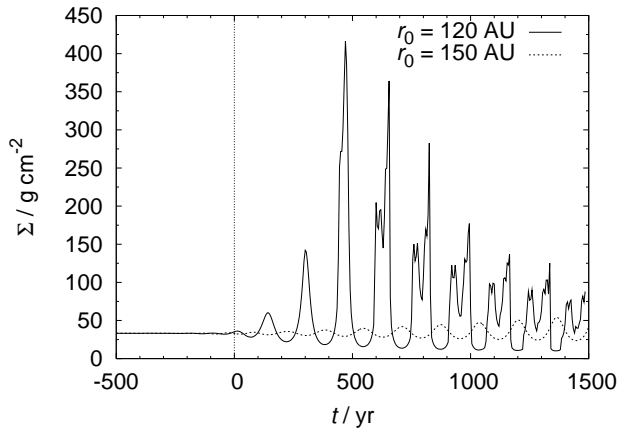}
\caption{\label{dens_120150au}Surface density $\Sigma$ vs time for
\mmbox{\rperi=120} and 150~AU measured in one probe
at a radius of 30~AU. In the closer fly-by $\Sigma$ reaches easily
10~times the initial value about 500~yr after periastron. The wider fly-by leads only
to slight density variations, which illustrates the strong dependence on $\rperi$.
The peaks in $\Sigma$ are due to the density waves moving with a different velocity
than the Keplerian orbital motion.}
\end{center}\end{figure}

A close encounter of a 0.5~\tmsun\ star at 120~AU with $e=1.5$ gives a good illustration of
the fundamental effects of tidal perturbations on a circumstellar disk.
Fig. \ref{six_120au} shows the overall development of the perturbed disk within 50~AU
during and after an encounter at 120~AU periastron distance.
Within only 400 years strong density
fluctuations appear as thin tidal arms of highly compressed material.
These tidal arms persist a few $10^2$ years up to $10^3$ years,
depending on the mass and the encounter distance of the perturber.
Within the spiral arms, material can be carried far outwards and even become unbound.
The time-dependent surface density is shown in Fig. \ref{dens_120150au} for encounter
distances of 120~AU and, for comparison, 150~AU. It can be seen that
the fly-by distance has a strong influence on the magnitude of the $\Sigma$ peaks; while
a 120~AU fly-by causes an increase of more than a factor of 10, the 150~AU encounter does
little more than doubling $\Sigma$.

\begin{figure}\begin{center}
\includegraphics[width=0.47\textwidth]{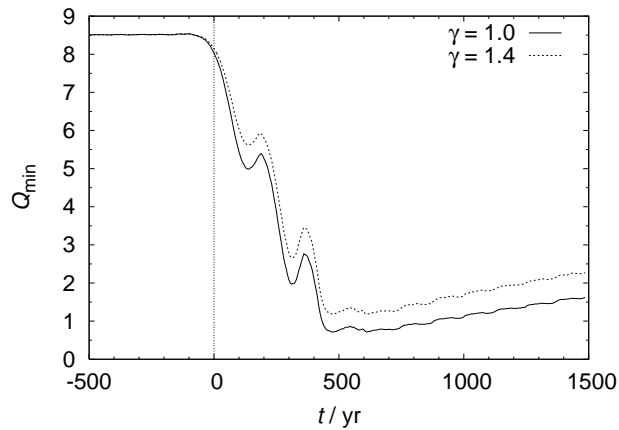}
\caption{\label{qmin_120au_ia}%
Minimum values of $Q$ within the region \mmbox{25~AU\le r \le 30~AU} (initial conditions)
during a 120~AU encounter of a 0.5~\tmsun\ perturber for the isothermal and the adiabatic
case (\mmbox{\gamma=1.0} and \mmbox{\gamma=1.4}, respectively). Even in the adiabatic
case $Q$ grazes the zone of Toomre instability
(\mmbox{\qmin\approx 1.7}), whereas the more realistic isothermal case almost
ensures the Toomre instability criterion (\mmbox{\qmin<1}).}
\end{center}\end{figure}

As the fly-by proceeds we follow $Q$ in each counting volume and at each time step the minimum
$Q$ is selected.
A plot of the $\qmin$ values within the computed disk area is shown in Fig. \ref{qmin_120au_ia}
for the isothermal (adiabatic index \mmbox{\gamma=1}) and the $\mathrm{H_2/He}$ adiabatic
case (\mmbox{\gamma=1.4}).
A disk with a locally isothermal equation of state and an initial \mmbox{Q=8.5}
exhibits regions with \mmbox{Q<0.8} within less than 500~yr after a perihelion passage closer
than about 120~AU, satisfying the Toomre instability criterion ($Q<1$). For
the adiabatic border case, $Q$ locally falls to 1.2, still below the upper limit for Toomre
instability \mmbox{Q\approx1.7}.
According to \citet{2004ApJ...610..456B}, the combined convective and radiative cooling is sufficient
to form clumps similarly to the isothermal case. Cooling times are of the order the orbital period
even for optically thick disks and there fulfil the criterion for fragmentation investigated by
\citet{2001ApJ...553..174G}.
Thus, an isothermal equation of state is expected to be a reasonable approximation.

\section{Results of the parameter study}
\label{sec:parscan}
We surveyed the parameter space \mmbox{(\mstar,e,\rperi)}
in order to find the regions where $Q<1$ locally occurs.
The parameters have been varied successively with appropriate step sizes
(Tab. \ref{tab:parrange}).

Here we describe the major results of our parameter study dealing with
the triggering of gravitational instabilities, the masses and orbits of seed candidates for
coplanar prograde passages, and the angular momentum tilt due to inclined passages.
The possibility of triggering for inclined passages has been studied for isolated cases.

\begin{table}\begin{center}
\begin{tabular}{cccc}\hline
        &from        &to          &with step\\\hline
$\mstar$&$0.1\,\msun$&$1.0\,\msun$&0.1--0.2~$\msun$\\
$e$     &1.0         &10          &0.5--1.0\\
$\rperi$&60~AU       &200~AU      &5~AU\\\hline
\end{tabular}
\caption{\label{tab:parrange}Ranges and step sizes for the varied parameters
$\mstar$, $e$ and $\rperi$}
\end{center}\end{table}

\subsection{Triggered Toomre instability for prograde passages}
\label{ssec:parscantoomre}

\begin{figure}\begin{center}
\includegraphics[width=0.49\textwidth,bb=83 75 413 255]{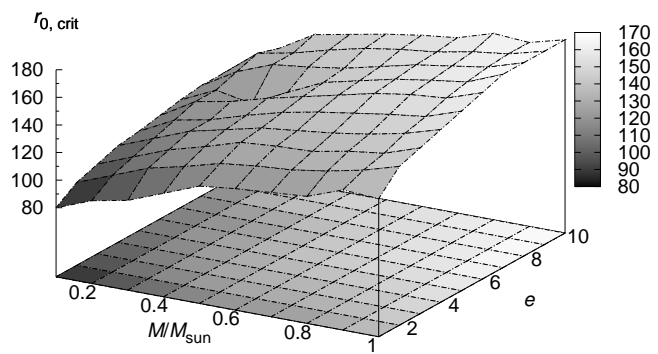}
\caption{\label{isdvt}Upper border $\rkrit$ of the regime of $Q<1$ occurrence in the
$(\mstar,e,\rperi)$ parameter space for the isothermal equation of state.
$Q<1$ occurs for values of $\rperi$ below the border indicated
by the grid surface. Clearly visible is the increasing $\rperi$ trend for increasing perturber mass
and eccentricity. The latter trend is caused by the decreasing relative angular velocity between
perturber and test particles. The key on the right indicates the $\rperi$-contour levels
plotted in the $m,e$-plane.}
\end{center}\end{figure}

Our parameter study for prograde coplanar passages shows that the Toomre instability
criterion \mmbox{Q<1} is locally satisfied for all stellar masses above 0.1~\tmsun\
and all eccentricities between 1 and 10 for encounter distances
\mmbox{\rperi\le80\unit{AU}}. For perturber stars with 1~\tmsun\ and $e=10$ we found $Q<1$
occurrence even for fly-bys with \mmbox{\rperi\le170\unit{AU}}.
This critical encounter radius $\rkrit$, the upper limit for
$\rperi$ up to which the Toomre instability criterion is satisfied,
depends on both the perturber mass and the encounter eccentricity in a monotonically
increasing manner, as can be seen in Fig. \ref{isdvt}. This reveals two major trends:
Increasing the mass leads to an increased maximum perihelion distance $\rkrit$, and also
a higher eccentricity of the perturber orbit results in larger $\rkrit$.

The mass dependency is simply caused by an enhanced perturbation with increasing mass.
The eccentricity dependence on the other hand is less obvious.
Unless the encounter velocity is extremely high or the periastron
distance extremely small (i.e. close to the particle orbits) the angular velocity
of the perturber is normally significantly smaller than that of the disk particles.
Thus an increment of the encounter velocity for a given
periastron radius yields a larger $e$ and a smaller particle-perturber relative velocity, i.e.
longer exposure time.

\begin{table}\begin{center}
\begin{tabular}{|c|cccccc|}\hline
&\multicolumn{6}{|c|}{Eccentricity $e$}\\
\raisebox{1.5ex}[-1.5ex]{$\mstar/\msun$}&1&1.5&3&5&10&50\\\hline
0.1&0.77&0.71&0.61&0.53&0.44&0.26\\
0.3&0.73&0.68&0.58&0.50&0.41&0.25\\
0.5&0.69&0.64&0.55&0.48&0.39&0.24\\
0.7&0.67&0.62&0.53&0.46&0.38&0.23\\
1.0&0.63&0.58&0.50&0.44&0.36&0.21\\\hline%
\end{tabular}
\caption{\label{tabresonanz}Coupling radius $a_0$ in units of $\rperi$, depending on the mass and orbit of the perturber}%
\end{center}\end{table}

An important question is how much mass is included in a typical probe with observed
\mmbox{Q<1}. A typical probe in our simulation has a radius \mmbox{h=0.1\unit{AU}} (Appendix
\ref{sec:denseval}), but the total region of a local Toomre instability could be larger
than that. In an unperturbed disk $\Sigma$ has a value of about \mmbox{40\unit{g\,cm^{-2}}} or
\mmbox{1.5\,\mterr\,\mathrm{AU}^{-2}} at 30~AU, which results in about 0.03 Earth masses
($\mterr$) within $h$. Within a tidal arm $\Sigma$ can reach 10 to 20 times this value,
yielding 0.3 to 0.6~$\mterr$ per probe.

Since the probe contains only a fraction of the Toomre-unstable material of the tidal arm,
the total mass of the unstable region is likely to be much larger. A typical compressed
arm is about 0.5~AU wide. A spherical region of 0.5~AU contains 10--20~$\mterr$ or about
one Neptune mass. If one per cent of this is dust \citep{Natta04} a body of about
0.1--0.2~$\mterr$ can form out of this seed. This would skip much of the coagulation time
otherwise needed, and is about 100 times more massive than the most massive KBOs.

\subsection{Triggered Toomre instability for retrograde and inclined passages}
\label{ssec:retrotoomre}
\begin{figure}\begin{center}
\includegraphics[width=0.47\textwidth]{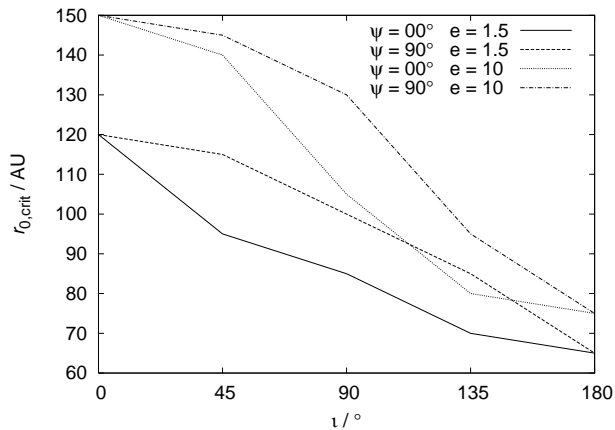}
\caption{\label{incencs}Critical upper pericentre distance $\rkrit$ for a sample of encounters
with non-zero inclination $\inc$ compared to the corresponding coplanar prograde fly-by.
The perturber mass is 0.5~\tmsun in all cases. Retrograde passages result in $\rkrit$ about
50\pct\ of that for zero inclination independently of the eccentricity. $\psi$ is the angle
between the pericentre and ascending node.}
\end{center}\end{figure}

In a few examples we also studied the triggering effects of an inclined prograde
and a retrograde passage
of a 0.5~\tmsun\ perturber. We varied the inclinations between 45\degr\ and 180\degr\
in steps of 45\degr.
As one would expect, the perturbation in both cases is smaller than that of a
coplanar prograde passage being due to the larger distance between the perturber and that
edge of the disk with the lowest angular velocity relative to the perturber.
The results are summarised in Fig. \ref{incencs}.

For the coplanar retrograde passage (inclination = $180\degr$),
we found \mmbox{\rkrit=65\unit{AU}} for \mmbox{e=1.5} and
\mmbox{\rkrit=75\unit{AU}} for \mmbox{e=10}. Both are about half the $\rkrit$ in the coplanar
prograde case,
i.e. the reduction of $\rkrit$ due to inclination is apparently independent of the eccentricity.
For other inclinations the decrease of $\rkrit$ depends on the angle $\psi$ between the pericentre
and the ascending node, where \mmbox{\psi=90\degr} leads to significantly higher $\rkrit$ than
\mmbox{\psi=0}.

\subsection{Tilt of the angular momentum vector}
\label{ssec:tilt}
\begin{figure}\begin{center}
\includegraphics[width=0.47\textwidth]{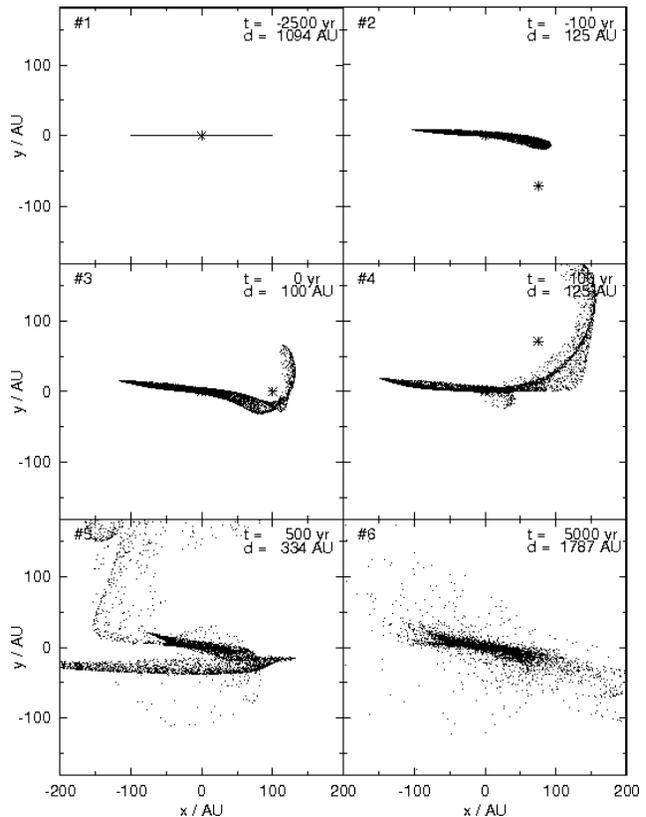}
\caption{\label{six_tilt}Parabolic fly-by of a 0.5~\tmsun\ star at 100~AU with 45\degr\
inclination. The angle $\psi$ between pericentre and ascending node is $90\degr$.}
\end{center}\end{figure}

In inclined encounters there is the possibility for tilting the plane of the disk due
to the transfer of angular momentum to the orbit of the perturber. We investigate the
possible origin of the obliquity of the Sun's rotational axis relative to the ecliptic
for inclinations $\inc$ between 5\degr\ and 175\degr and \mmbox{\psi=0}.
Passages with inclination less than 90\degr\
are designated as prograde, and those beyond 90\degr\ as retrograde. In this section the
radius of the model disk is set to 100~AU to measure the angular momentum
change of all parts of the disk, because we expect the resulting tilt to be close to
the mean tilt of the particle orbits and a global rather than a local effect.
The global effects
on the disk are shown in Fig. \ref{six_tilt}.

\begin{figure}\begin{center}
\includegraphics[width=8.4cm]{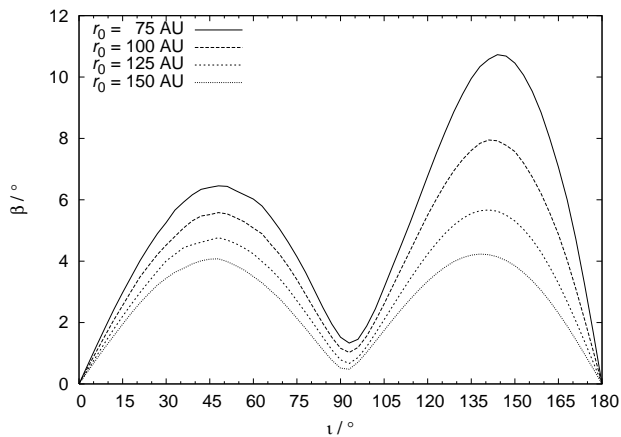}
\caption{\label{tilt_a090}Tilt $\tilt$ of the disk after fly-by of a 0.5~$\msun$ perturber
as a function of inclination and closest distance. $\inc<90\degr$ are prograde encounters,
$\inc>90\degr$ retrograde ones.}
\end{center}\end{figure}

After the encounter, when the distance between the
stars is large enough to prevent significant further tidal effects, the sum of the orbital
angular momenta of all disk particles is calculated. 
The results are plotted in Fig. \ref{tilt_a090}: the most remarkable feature is that
the tilt has its maximum of 13\degr\ for \emph{retrograde} passages of about 140\degr\ and
\mmbox{\rperi=75\unit{AU}}, whereas
the local maximum in the prograde area at 45\degr\ appears much lower for close
encounters below 100~AU (6.5\degr\ for 75~AU). For less close encounters, however, the maximum tilts
are essentially equal for both prograde and retrograde passages.
In both cases, an encounter below about 125~AU results in a maximum tilt
of more than 6\degr\, which is close to the observed solar tilt of 7\degr.
We return to this point in \S~\ref{ssec:hellersph}.

\section{Discussion}
\label{sec:discussion}
\subsection{Cooling and fragmentation of disks}
Based on an analytical estimate \citet{Rafikov2005} found that 
the fragmentation of protoplanetary disks requires unusual high surface densities
and temperatures.
According to his paper a local Toomre instability does not necessarily
lead to fragmentation. It is crucial that the instability can be achieved \emph{despite}
temperatures which are high enough to provide efficient radiative cooling via the Stefan-Boltzmann
law. According to \citet{Rafikov2005} the temperature $T_\mathrm{crit}$ required for efficient cooling at 10 AU ranges between 200 and more than 1000 K depending on the disk model.
Only if the disk mass is a factor of 10 higher than that used by \citet{2004ApJ...610..456B}
is a Toomre instability possible. Because, even if the initial temperature of the condensed region
is less than $T_\mathrm{crit}$, the cloud being more unstable, the rapid increase of
temperature during gravitational contraction eventually compensates the increasing pressure.

The minimum disk surface density $\Sigma_\mathrm{min}$ and the critical temperature
$T_\mathrm{crit}$ are given by
\citep{Rafikov2005}
\begin{equation}\label{raf05S}
\Sigma\ge\Sigma_\mathrm{min}
=\Omega^{7/5}(\pi G Q_0)^{-6/5}\left[\frac{1}{\zeta\sigma}\left(\frac{k}{\mu}\right)^4\right]^{1/5}
\left[f(\tau)\right]^{1/5}
\end{equation}
and
\begin{equation}\label{raf05T}
T\ge T_\mathrm{crit}
=\Omega^{4/5}\left(\zeta\pi Q_0 G\sigma\right)^{-2/5}\left(\frac{k}{\mu}\right)^{3/5}
\left[f(\tau)\right]^{2/5}\fkomma
\end{equation}
where $Q_0\approx1$ is the upper Toomre limit for disk instability, $\zeta=2\xi(\gamma-1)$ 
is a parameter to be satisfied by the cooling time $t_\mathrm{cool}$ and the orbital
frequency $\Omega$, \mmbox{\Omega t_\mathrm{cool}<\xi\sim1}.
Note that our $T_\mathrm{crit}$ equals $T_\mathrm{min}$ in \citet{Rafikov2005}.
$\sigma$ is the Stefan-Boltzmann constant, $k$ the
Boltzmann constant, \mmbox{\mu\approx2.3\unit{u}} the molecular mass of the disk gas and
\begin{equation}
f(\tau)=\tau+\frac{1}{\tau}\fkomma
\end{equation}
where $\tau$ is the optical depth which we estimate to be in the order of several tens in a
compressed region.
The value of $\tau$ depends on the opacity and the
surface density of the particular disk region.
Given the model parameters used in this paper, an estimate of \mmbox{\tau=60} of the compressed
disk region and \mmbox{\xi=1},
conditions (\ref{raf05S}) and (\ref{raf05T}) require \mmbox{\Sigma\ge630\unit{g\,cm^{-2}}}
with \mmbox{T_\mathrm{crit}=330\unit{K}} at 30~AU. Numerical simulations suggest \mmbox{\xi\approx3}
\citep{2001ApJ...553..174G}; in that case the limits are \mmbox{\Sigma_\mathrm{min}=510\unit{g\,cm^{-2}}}
and \mmbox{T_\mathrm{crit}=210\unit{K}}. Assuming a tidal compression by a factor of 15 (which is
a reasonable value in our simulations) the surface density condition is easily achieved in the
compressed area, even with significantly lower disk masses than proposed by \citet{2004ApJ...610..456B}
or \citet{2004ApJ...609.1045M}.
The assumed optical thickness in these cases corresponds to a mean
opacity \mmbox{\omega=\tau/\Sigma\approx0.1\unit{cm^2\,g^{-1}}} of the disk material,
which is a reasonable value for
condensed regions in protoplanetary disks (\citealt{Podolak2004};
\citealt{PKC85}, for centimetre-sized dust grains).

Another problem which we do not treat in this paper
is the transformation of the presumably unstable regions into 
(proto-)planets. In case of the formation of Neptune or Uranus this
includes the question of how to get rid of a large amount of gas 
with respect to the dust, by this ending up with the observed high
mass fraction in solid material in these planets.

In principle, the following scenarios for the evolution of a gas-rich
dust disk are possible:

A dust phase can partly decouple from the gas, resulting in a comparatively
thin dust layer (\citealt{Natta04}; \citealt{BFetal2005}).
Since the dust evolves basically pressure-free it is much more prone to 
gravitational instability than the gas and, hence, instabilities
in the dust phase could grow faster and form denser regions than the
corresponding instabilities in the dynamically hotter gas.
A plausible way of increasing the global solid-to-gas ratio is as follows:
Jupiter and Saturn form traditionally within a few Myr. Meanwhile the disk
is partially photo-evaporated such that the dust-to-gas ratio increases
significantly, at which time the fly-by-perturbation occurs. From Fig.
\ref{isdvt} we see that gravitational instabilities can be induced for
\mmbox{\rkrit\le200\unit{AU}}, and from Fig. \ref{penc} it can be seen that
the probability of such an encounter lies between 10 and 60\pct\ if the
Sun was born
in a cluster with a radius of $\sim$0.5~pc containing
$1000\le N\le10000$ low-mass stars (one encounter within $\sim$10~Myr).
In this scenario the GIs would be induced in a dust-dominated disk
(see \S~\ref{ssec:diskprop}).

A preferential loss of gas locally might also be possible
when the clumps survive sufficiently long and mass segregation by gravity
brings the dust to the center of the clumps, by this shielding the
dust from a later loss by photo-evaporation. Boss (2004) suggested that 
the clumps can survive for a few dynamical periods which might be 
sufficient for the required mass segregation. Moreover, mass segregation
can be possibly sped-up by vortices, as \citet{KlahrBodenheimer2004} show.
The solid component tends to settle rapidly in the center of a vortex due
to dynamical friction. \citet{KlahrBodenheimer2003} describe the formation of
vortices from baroclinic instabilities. However, we expect vortices to occur
also in gravitational instabilities if there is an initial spin angular momentum.
The contracting GI will then be spun-up by the Coriolis forces. 

In a border case the gas fraction
may have been lowered due to normal depletion mechanisms during disk
evolution. In the case that the dust may have already settled in or near the mid-plane,
the horizontal centring due to vortices will result in an extreme concentration
of dust within a small volume, thus skipping the otherwise necessary vertical
contraction due to self-gravity.

  We want to stress that the discussed possibilities for getting high 
dust mass fractions in the clumps (i.e.\ an initial value problem or 
an evolutionary process) cannot be decided on 
the basis of our simple simulations. A more refined multi-component 
treatment of dust and gas as separate phases (e.g.\ as done
by \citet{Theis&Orlova2004} in the case of dusty galactic disks)
would allow for investigating these in detail, but this would be
part of a future investigation. We do believe, however, that our found
unstable regions would be excellent  large-scale seeds acting as birth
places of (proto-)planets.

\subsection{Orbits of candidate planetesimals}
\label{ssec:exorbs}
On a more speculative note our simulations allow us to study the likely initial shapes of the
orbits of candidate protoplanetary
seeds, which are chosen here as the probes where \mmbox{Q<1}, indicating the possible
formation of a planetary core.
The flyby distance is 100~AU, and two cases, a coplanar and an
inclined orbit (\mmbox{\inc=45\degr}) are investigated.
The perturbations have their greatest impact on the region beyond 30~AU, i.e. the Kuiper belt region.

\begin{figure}\begin{center}
\includegraphics[width=8.4cm]{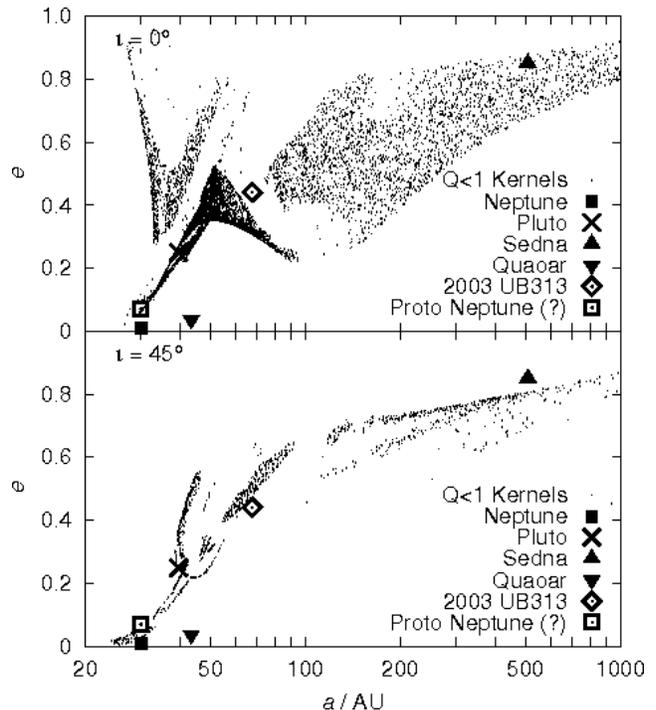}
\caption{\label{rqe}%
Distribution of semimajor axes $a$ and eccentricities $e$ of the \mmbox{Q<1}
probes (indicating candidate planetary seeds) after a parabolic fly-by at 100~AU with an inclination
of \mmbox{\inc=0\degr} (upper panel) and \mmbox{\inc=45\degr} (lower panel). The orbits of
some actual major bodies are included as well. Pluto, Sedna and 2003 UB313 are contained in or close
to the main population, whereas Quaoar lies outside and Neptune below the ``trail'' at 25--30~AU.
If the early Neptune had a slightly more eccentric orbit (\mmbox{e=0.07}) it would have fit
this plot.}
\end{center}\end{figure}

The distribution of the eccentricities $e$ and semimajor axes $a$ is shown in Fig.
\ref{rqe}, where from 10000 test volumes initially at radii between 20 and 100~AU
only those are plotted, for which \mmbox{Q<1} occurred at least once.

It can be seen
that the orbit of the KBO \emph{Sedna} is not far outside the bulk of candidate seeds,
and well contained in a less-dense peripheral region of candidate seeds for the coplanar
case. It is still close to it in an 45$\degr$ encounter. Also Pluto is
contained in this population, despite its resonance with Neptune, which most probably is the
result of orbital evolution after the formation. Neptune, however, lies below the candidate seed
population but close to a trail of low-eccentric particles. Also due to its low eccentricity
Quaoar lies outside the population.
The eccentricities in this plot range from 0.02 (nearly circular) to
1 (parabolic ejection) and would even exceed 1 if plotted over the perihelion distance rather than
the semimajor axis.
Even the recently announced large KBO 2003 UB313 fits this plot. However, its
the longitude of its ascending node deviates more from those of the other shown objects.
Note that, {\it if the present proposition is realistic, we expect our solar system
to be void of KBOs below and above the region populated in Fig. \ref{rqe}.}

We note that \citet{ida&larwood&burkert:2000}
propose that the eccentric orbits of the Kuiper-Belt objects may have formed as a result of tidal
deformation of a pre-existing planetesimal disk. Our alternative notion is that the eccentric
outer-solar-system objects may be a result of an encounter which lead to the formation of Uranus,
Neptune and the KBOs.

\subsection{Earlier SPH simulations of tilted disks}
\label{ssec:hellersph}
A more extensive study of the disk tilt due to stellar encounters than ours (\S~\ref{ssec:tilt})
has been made by Heller (1993),
who used an SPH model with disk-star and disk-perturber interaction instead of restricted
$N$-body computation. His results are essentially the same as in our study. Minor
differences due to the absence of viscosity are possible. According to Heller (1993),
the ability of the disk to deform, which is maximal for non-interacting test particles,
enhances the resulting disk tilt. The perturbation of the perturber-star motion
due to disk-star interactions also alters the results, but since the effects are small
in most cases \citep{heller:1993}, they can be neglected here. The important factor
causing Heller's results on the tilt to be somewhat larger than ours is due to his large
particle removal radius of 800~AU, whereas we neglected particles with a
semimajor axis larger than 150~AU.

\subsection{Limits and perspectives}
The restricted $N$-body method has its great advantage in low computational efforts. This allows us
to perform a wide parameter study within an acceptable computing time, but on the other hand this
also limits the accuracy of the answers.

Since we do not model self-gravity our approach is only reasonable if the mass of the disk is small
compared to the solar mass, making the latter the dominant source of gravity. The same applies for
the dynamical feedback on the orbit of the stars and the wobbling of the Sun (which would alter
the density evolution of the disk) due to asymmetric disk perturbations.

The absence of hydrodynamical interactions limits the simulation time within which reasonable results
for the density evolution can be expected. Furthermore, the effect of cooling mechanisms as
mentioned by \citet{2002ApJ...576..462B} and \citet{2004ApJ...610..456B} can only be approximated
by adjusting the adiabatic exponent $\gamma$. 

Due to these restrictions this work provides only a rough estimate of the expected magnitude of
triggered instabilities.
The evolution of the strong initial perturbations is determined mainly kinematically.
So the results are reliable if only the immediate effects are of interest. As time proceeds the
evolution will be more and more determined by hydrodynamical and other interactions within the disk.
But this uncertainty mainly affects the borders of the candidate parameter subset rather than the
gravitational-triggering scenario as a whole. 

Despite these limitations, our study gives us an important guideline for a more tightly focussed but
computationally extremely expensive hydrodynamic research.

\section{Conclusions}
\label{sec:concl}
Using a simple, but fast numerical method a parameter study has been made to test
the possibility of gravitationally triggered planet formation.
This scenario is supported by analytical estimates of the
relatively high encounter probability in a typical young cluster (Fig. \ref{penc})
as well as by the observed obliquity of the solar rotational axis being a possible
consequence of an inclined fly-by.

As the main outcome of this study we summarise the following results:
\begin{enumerate}
        \item Prograde passages of stars with masses of 0.1--1~\tmsun\ can trigger Toomre
              instabilities within a 30~AU radius in a disk of 0.07~\tmsun\ if an approximately
              isothermal behaviour of compression is assumed. Toomre instabilities are also
              expected for adiabatic compression. Higher perturber masses and eccentricities
              increase the density perturbations and therefore allow for larger areas to become
              unstable.
        \item Retrograde passages cause local $Q<1$ for smaller $\rperi$ (about half the value of
              the prograde case). Inclined passages yield $Q<1$ for
              intermediately close fly-bys.
        \item A fly-by of a 0.5~\tmsun\ star with an inclination between $30\degr$ and $60\degr$
              and \mmbox{\rperi\le100\unit{AU}} tilts the disk plane by $3\degr$--$6\degr$.
              This is very similar to the observed solar obliquity.
\end{enumerate}

The duration of the locally triggered gravitational instability is in the range of centuries to
millennia. Therefore, if these Toomre instabilities can lead to fragmentation and enhanced
core-formation they may solve the discrepancy between the short disk lifetime and the long
(classical) formation times of the outer planets and maybe even the Kuiper-Belt objects (KBOs).
Even if only a fraction of today's Neptune mass was assembled by triggering, this might be
the required seed for the ongoing coagulation and accretion to be finished before the disk
dissolves.
The high solid-to-gas ratio of Uranus and Neptune may be explained in this
scenario if their formation was induced by a fly-by at a solar-system age \mmbox{\ge5\unit{Myr}}
when the disk was depleted of gas. In this situation $Q_\mathrm{dust}<Q$ such that the dust disk
should be even more unstable towards GIs than suggested above.
This sets possible constraints on the physical properties of the Sun's birth cluster.

Our assumed disk mass is comparable to the minimum solar nebula mass. This avoids
implausible assumptions on the disk mass as they are required for pure gravitational fragmentation
of unperturbed disks.

\section*{Acknowledgements}
Pavel Kroupa acknowledges support through DFG grant 1635/4--1. We thank Simon Goodwin for very
useful discussions.



\begin{appendix}
\section{Density evaluation}
\label{sec:denseval}
Calculation of the local Toomre disk stability requires the
evaluation of the local surface density. At first, all disk
particles are weighted by an individual mass $m_b$ which
corresponds to its initial orbit radius $r_b$ in such a way that
\begin{equation}\label{summass}\sum_1^N m_b=\mdisk(r_1\le r\le r_2)\fkomma\end{equation}
where $r_1$ is the inner and $r_2$ the outer border of the simulated
disk region. Since the $N$ particles are initially positioned on a polar grid
consisting of $n_r$ rings with $n_\varphi$ particles each, which results
in a $r^{-1}$ law of the surface density $\Sigma$, the mass-weighting
must be proportional to $r^{-1/2}$ to fit the needed
$r^{-3/2}$ law (Eq.\ref{sigmaprofile}). Together with eqn. (\ref{summass}) this leads to
\begin{equation}\label{particlemass}
m(r_b)=\frac{r_b^{-1/2}}{\sum\limits_{b=1}^N r_b^{-1/2}}
\cdot\frac{\sqrt{r_2}-\sqrt{r_1}}{\sqrt{\rmax}-\sqrt{\rmin}}\cdot\mdisk\fkomma\end{equation}
where $\rmin$ and $\rmax$ are the inner and outer radius of the \emph{real} disk.
Here we use $\rmin=3\unit{AU}$ and $\rmax=100\unit{AU}$.

The density evaluation is achieved by simply adding the $m_b$ within
``counting-volumes'' surrounding massless probe particles which move
along with the disk particles on ballistic (Keplerian) orbits. Within the
counting-volume the density is calculated in SPH-style using
\begin{equation}\label{densa}\gw{\varrho}=\sum_{b=1}^N m_b\,W(d_b,h)\fkomma\end{equation}
where \mmbox{d_b} is the distance of the $b$-th mass-weighted test particle
from the center of the probe particle. The \emph{kernel function} $W$
provides a smooth distance-weighting of the test particles for reduction of
numerical noise. In our model the compact spline kernel from
\citet{monaghan&lattanzio:1985} is used.
The choice of the kernel radius $h$ determines the mean number of particles $n_h$ within
the probe volume. For a random distribution, \mmbox{n_h\ge100} is required to keep the noise
low enough. A regular distribution, as we used here, shows appropriate results even for
\mmbox{n_h\approx10}. For a given (average) $n_h$ and disk population parameters $r_1$, $r_2$
and $N$ we get $h$ from
\begin{equation}
h=\sqrt{\frac{n_h\left(r_1+r_2\right)\left(r_2-r_1\right)}{N}\,.}
\end{equation}
Thus, typically, \mmbox{h=0.1\,\mathrm{AU}} for \mmbox{r_1=25\,\mathrm{AU}}, \mmbox{r_2=30\,\mathrm{AU}}
and \mmbox{N=250000}.

Since we want to identify unstable regions, i.e. regions where $Q<1$ locally,
we applied about 1000 of these test
volumes, which concentrate at the areas of the highest density.
Their dynamic behaviour is the same as that of the disk test particles.
Thus, a high spatial resolution at the areas of interest is automatically
provided.
\end{appendix}
\end{document}